\documentstyle[sao1,psfig]{article}
\def\sec{\ifmmode {}^{\prime\prime}\else ${}^{\prime\prime}$\fi~}

\def\magdot{\ifmmode {}^{\rm m}\!\!\!.\, \else ${}^{\rm m}\!\!\!.\,$\fi}
\def\asec{\ifmmode ^{\prime\prime}\else$^{\prime\prime}$\fi}
\begin{document}
\Large
\vspace*{-0.5cm}
\begin{center}
{\Large \bf Face--on SS\,433 stars as a possible new type of
extragalactic X--ray sources}\\
S.\,Fabrika\\
\vspace*{-.3cm}
Special Astrophysical Observatory, Nizhnij Arkhyz, Russia\\
\vspace*{-.3cm}
({\it email: fabrika@sao.ru})\\
 A.\,Mescheryakov\\
\vspace*{-.3cm}
Sternberg Astronomical Institute, Moscow, Russia \\
\vspace*{-.3cm}
({\it email: shura@sai.msu.ru})
\end{center}
\vspace*{-0.9cm}
\section*{Abstract}
\vspace*{-.5cm}
The SS\,433 objects is a well--known source of relativistic jets,
which are formed in supercritical accretion disk. It is very
probable that the disk has polar channels and their radiation
is collimated (the photocones). The face--on SS433 object can
appear as beamed ultra--bright (and highly variable) X--ray source,
$L_x \sim 10^{39} - 10^{42}$ erg/s. We discuss properties of these
hypothetical objects and their frequency expected
in galaxies. We describe a search for such objects using the ROSAT
ASS Bright Source Catalog and Faint Source Catalog
(99528 point--like sources) and the RC3 catalog (16741 spiral
and irregular galaxies). Among in total 418 positive correlations 
in all types of galaxies we find that in S and Irr galaxies 142 sources
are unknown as AGNs.
We isolated by visual inspection 37 clear non-nuclear, 43 probably
non--nuclear and 35 probably nuclear sources. The last two classes 
certainly contain many unknown AGNs. The sources of the 1-st class
have X--ray luminosities mainly $10^{39} - 3\,10^{41}$ erg/s. Their observed
frequency is about $4 \div 5\,\%$ per S/Irr galaxy,
what is in agreement with
expected frequency of face-on SS\,433 stars. The only way to recognize
such stars is their expected violent variability in X--rays.

\section*{Backgrouds}

The well--known galactic binary star SS\,433 shows unique relativistic jets
moving with a velocity $V_j = 0.26\,c$.
The jets are originated in the supercritical accretion disk.
The disk is extremely bright, its total luminosity is
$L_{bol} \sim 10^{40}$~erg/s with maximum of the radiation in UV
(Dolan et al.~1997). We observe a supercritical wind outflowing from
the disk, the temperature of the wind photosphere is
$T_{ph} \approx 5 \cdot 10^4$~K. The disk orientation is ``edge--on''
and variable ($\pm 20^{\circ}$) because of the precession.

The jets propagating in inner regions of the wind make channels,
which may form a collimated radiation of SS\,433. The jets are
accelerated and collimated in some inner region inside the photosphere.
They are accelerated by radiative pressure with the line--locking, where
the collimated radiation is also needed (Shapiro, Milgrom, Rees~1986).
Necessity of SS\,433 collimated radiation follows also from observation
data (Fabrika~1997). The outflowing wind velocity is $V_w \approx 1500$~km/s,
the total mass loss rate in the wind is
$\dot M_w \sim 10^{-4}\,M_{\sun}/y$, the mass loss rate in the jets
is $\dot M_w \sim 5 \cdot 10^{-7}\,M_{\sun}/y$. The wind photosphere size
is $R_{ph} = \dot M_w \sigma_T /4 \pi m_p V_w
\approx (1 \div 2) \cdot 10^{12}$~cm and corresponding photospheric
temperature is  $T_{ph} \approx 5 \cdot 10^4$~K, what agree well with
direct estimates from SS\,433 fluxes.
The photospheric size inside the cannels
$R_{phj} = \dot M_j \sigma_T / \pi (\theta_c/2)^2  m_p V_j
\sim 10^{10}$~cm and corresponding temperature of inner photosphere
is $T_{ph} \sim 5 \cdot 10^5$~K, where $\theta_c$ is opening angle
of the channels. The wind photosphere mean size $R_{ph}$ is much
greater than that of around the jets $R_{phj}$, what means that
channels and collimated radiation do exist. From SS433 data we can only find
upper limit on the opening angle of the photocone $\theta_c /2 < < 60^{\circ}$.

The supercritical accretion disk simulations (Eggum, Coroniti, Katz,\,1985)
show that in the inner regions of accretion disk a cone is formed
in outflowing matter,
where the collimated radiation and fast--moving gas propagate. These
cones are rather broad, $40^{\circ} - 60^{\circ}$. In active galactic
nuclei (AGNs) ionization cones are observed, which are very probably
photocones
of collimated radiation. They are broad with opening angles in the range
$\theta_c = 40^{\circ} \div 100^{\circ}$ (Wilson, Tsvetanov~1994).

SS\,433 star being orientated ``face--on'' and observed in photocone
will look brighter by the geometrical brightening factor
$B = 2\pi/\Omega_c$, where $\Theta_c$ is solid angle of the channel.
It will be X--ray source
registered with a luminosity $B D^{3+\alpha} L_c$,
where $L_c$~-- the cone luminosity, $D=(1-(V_j/c)^2)^{1/2}/(1-V_j/c)$~--
the Doppler factor, $\alpha$~-- spectral index ($I\propto\nu^{-\alpha}$).
Accepting $\alpha = 1$ we find for SS\,433 the relativistic beaming factor
$D^{3+\alpha} = 2.9$. Ii is not easy to estimate $L_c$.
We may expect $L_c \sim L_{bol}$ and accept here $L_c = L_{bol}$.

How many such stars one may expect? We know one SS\,433 in the Galaxy
as a persistent superaccretor and a source of relativistic jets.
Results of relativistic binary stars
population synthesis (Lipunov et al.~1996) show that we may expect
($0.1 \div 10$) SS\,433s per a galaxy like Milky Way. Very important, that
it is much more propable to find such a star in young star--burst regions,
i.~e. rather in nuclear regions in galaxies. Let us accept one SS\,433 star
per spiral galaxy like MW, and taking into account the specific orientation
of ``face--on SS\,433'', we accept a frequency of the face--on SS\,433s
as $1/B$ per spiral galaxy.
So we have {\it an assumption} which is not easy to be corrected today~---
the cone luminosity is $L_c \sim L_{bol} = 10^{40}$~erg/s. We do not specify
opening angle $\theta_c$ (t.~e. the brightening factor $B$) of the photocone.
In Fig.\,1 we show these simple relations, how
expected X--ray luminosity ($L_x =B L_c $) and number of MW--like galaxies per
one face--on SS\,433 star ($1/B$) depend on $\theta_c$.

\begin{figure}
\centerline{\psfig{figure=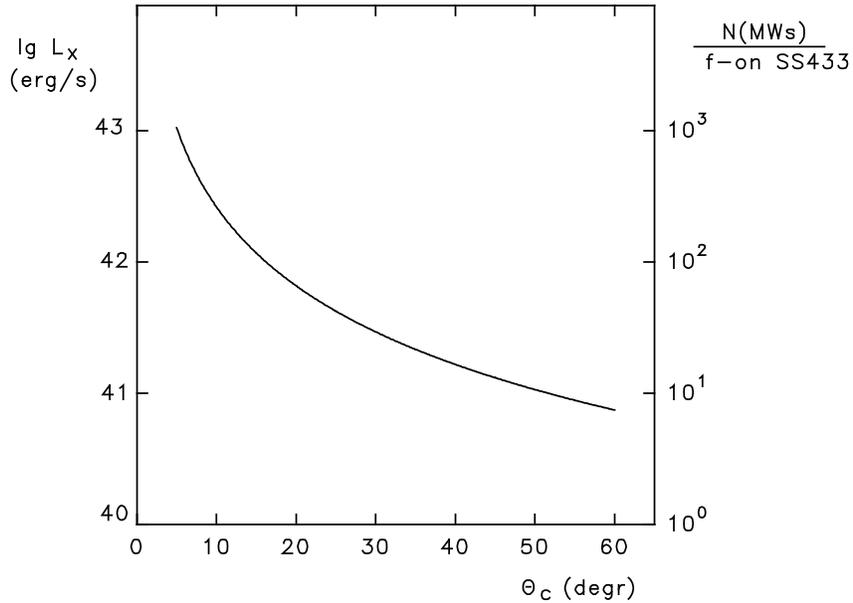,height=8cm}}
\caption{Expected X--ray luminosity ($L_x =B L_c $) and a number of
MW--like galaxies per one face--on SS\,433 star as depending on a cone
opening angle}
\end{figure}
\vspace*{-0.5cm}
\section*{Method}
\vspace*{-0.5cm}
We search for {\bf hypothetical} objects, face--on SS\,433 stars, which
have to be ultraluminous X--ray sources. Their expected
properties are:\\
{\bf Frequency}.
$1/B$ per spiral galaxy like MW. Th expected scatter is $\pm$~one order
of magnitude.\\
{\bf Location}.
Spiral galaxies: arms, star--burst regions, Irr galaxies. Nuclear regions
of spiral and S0 galaxies: the conditions needed for production of such
stars are practically the same as those for production of an AGN.\\
{\bf Luminosity and spectrum}.
{\it X-rays}: luminosity is ultra--bright, $10^{39} \div 10^{42}$~erg/s.
It depends on many unknown parameters: the cone luminosity and opening
angle, specific orientation and precession phase at observations.
Spectrum is unknown, probably soft. Very variable, sporadical variability
from 1~min to 1 year, periodical variavility with orbibal and
precession periods (days, months). The fast variability is the main criterion
for such objects. It could be quasi--periodical with a time--scale of
about  $\delta t \sim R_{ph}/V_j$, what is about 200~sec for SS\,433.\\
{\it Visible range}: very faint (Fabrika, Sholukhova 1995),
$V > 21^m$ for $D= 10$~Mpc. The main
contributor is the  accretion disk wind, the object appears as
UV or blue star.\\
{\it Radio}: the SS\,433 being face--on and
observed at $D= 10$~Mpc (with taking into account
relativistic beaming, $V=0.26\,c$) will show a radio flux
$4 \cdot 10^{-3}$~mJy. The transient source Cyg\,X--3 (accepting the same
beaming $V=0.3\,c$) will be observed as $5 \cdot 10^{-2}$~mJy source
in maximum. The transient source GRS\,1915+105 (the beaming factor
at $V=0.92\,c$) will show maximum flux of about $0.8$~mJy.
At the modern sensitivity of radio surveys, like the FIRST Survey we have
a little hope to detect such sources.

\begin{figure}
\vspace{-3cm}
\centerline{\psfig{figure=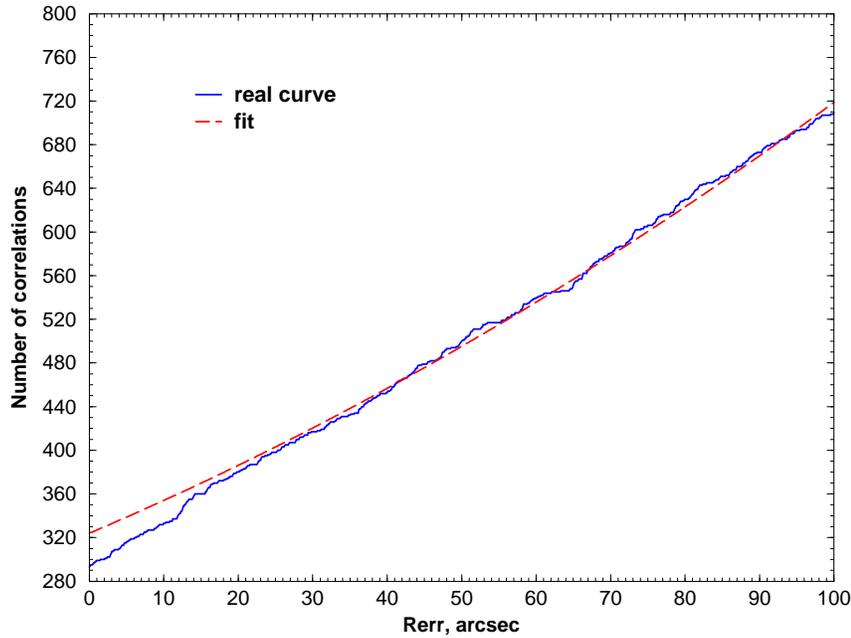,height=10cm,angle=-90}}
\caption{The number of positive correlations versus Rerr for BSC\,+\,FSC
sources. The polynomial fitting shows real sources + contanimations}
\end{figure}

We have correlated RC3 catalogue of galaxies (de Vaucouleurs et al.~1991)
with ROSAT All Sky Survey Bright Source Catalogue (BSC, Voges et al. 1999)
and Faint Source Catalogue (FCS, Voges et al. 2000). RC3 contains
23007 galaxies (15415 spirals and 1326 irregulars), it is considered
as complete up to $V = 14\magdot5 - 15\magdot0$. Distances to nearby 
galaxies ($<$\,5--10~Mpc) in RC3 may have large errors, we used data from 
Karachentsev et al.~(2000) to correct these distances. The Local Group
galaxies were omitted. ROSAT ASS  
(0.1~-- 2.4~kev) contains 18811 sources in BSC ($F > 0.05$~cts/s),
8365 among them are point--like and 91163 point--like sources in FSC. 

\begin{figure}
\vspace{-3cm}
\centerline{\psfig{figure=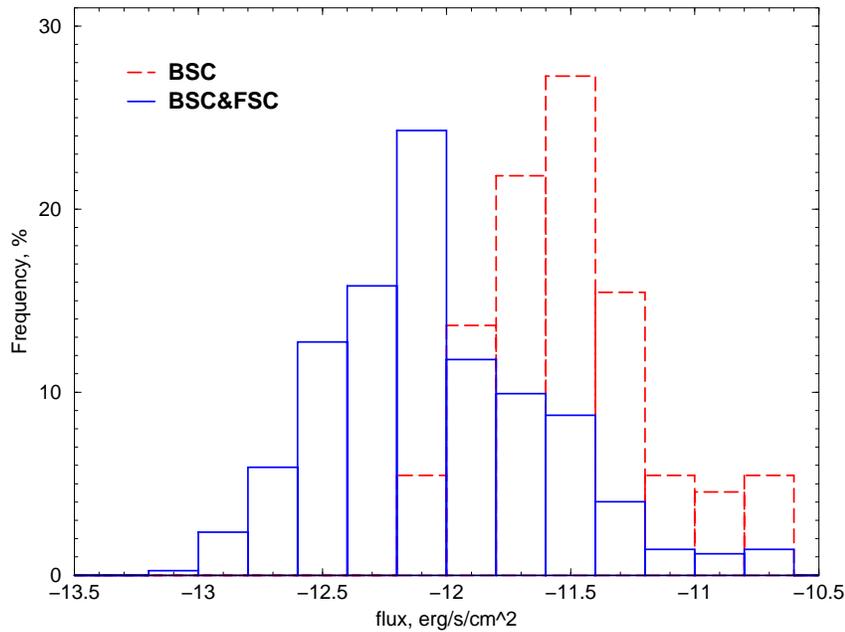,height=10cm,angle=-90}}
\caption{Frequencies of X--ray fluxes of all originally found
418 positive correlations}
\end{figure}

The correlation box was used as 25th magnitude isophote of a galaxy + Rerr.
Fig.\,2 shows the number of positive correlations versus Rerr together 
for the BSC and FSC. The polynomial fitting shows real sources + 
contanimations. The polinomial was found making multiple 
correlations with galaxy coordinates
$+ \Delta \alpha = 1^h, 2^h,$~etc. The deficit of galaxies below the
polinomial is contributed both by real errors in X--ray sources and 
galaxy coordinates and by extended sources. We have found that the
Rerr\,$<$\,30\sec and chosen 
Rerr~=~30\sec for final correlations. We see that about 400--420 
positive correlations do exist, but about 90 sources from them 
are contaminations (Fig.\,2).  
To select AGNs among our positive correlations we used
Veron--Cetty, Veron~(2000) cataloque choosing only confirmed
from optical spectroscopy Seyfert and Liner nuclei.
As a first step we study in this paper only point--like sources from ROSAT
ASS and only spiral and irregular galaxies from RC3 cataloque.
\vspace*{-0.5cm}
\section*{Results}
\vspace*{-0.5cm}
We have found 418 positive correlations of ROSAT ASS point--like sources~---
308 sources from FSC and 110 sources from BSC. After comparing with 
Veron--Cetty, Veron~(2000) cataloque we have 327 positive correlations
of X--ray sources (unknown as AGNs) with RC3 galaxies. Considering only
spiral and irregular galaxies, we find the final number of X--ray
sources is 142. 

Fig.\,3 shows frequencies of X--ray fluxes of all originally found
418 sources. The fluxes have been found accepting power--law spectrum
with the photon index
$\Gamma = 2.3$ and the Galaxy foreground absorption (Voges et al.~1999).
From this figure we find that FSC may be complete up to 
$6.3 \, 10^{-13}\,erg/cm^2\,s$. Only 10\,$\%$ of our sources were detected
in FSC below $2.7 \, 10^{-13}\,erg/cm^2\,s$. In BSC only 6\,$\%$ of our
positive correlations were detected below $10^{-12}\,erg/cm^2\,s$.

\begin{figure}
\vspace{-3cm}
\centerline{\psfig{figure=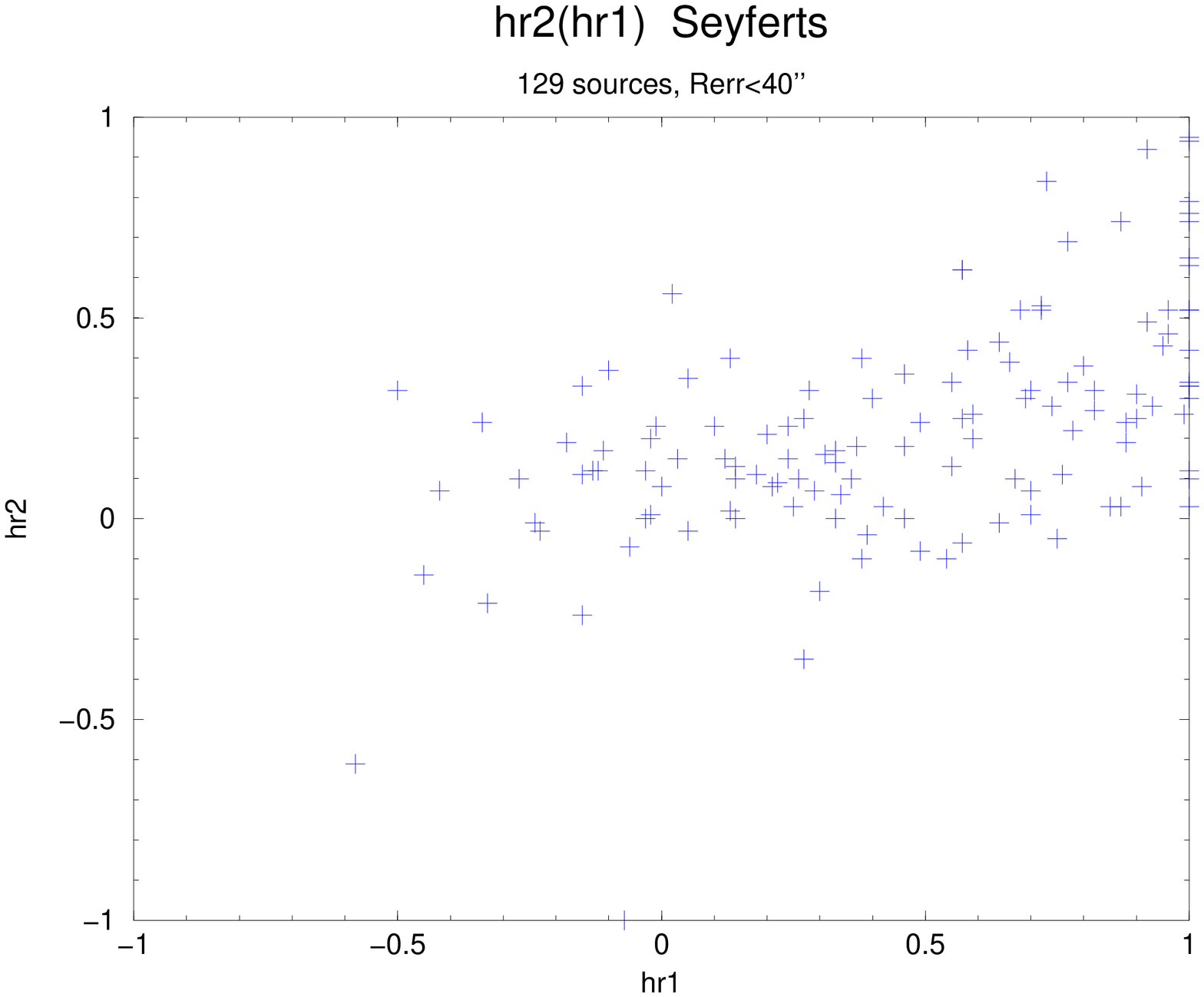,height=10cm,angle=-90}}
\centerline{\psfig{figure=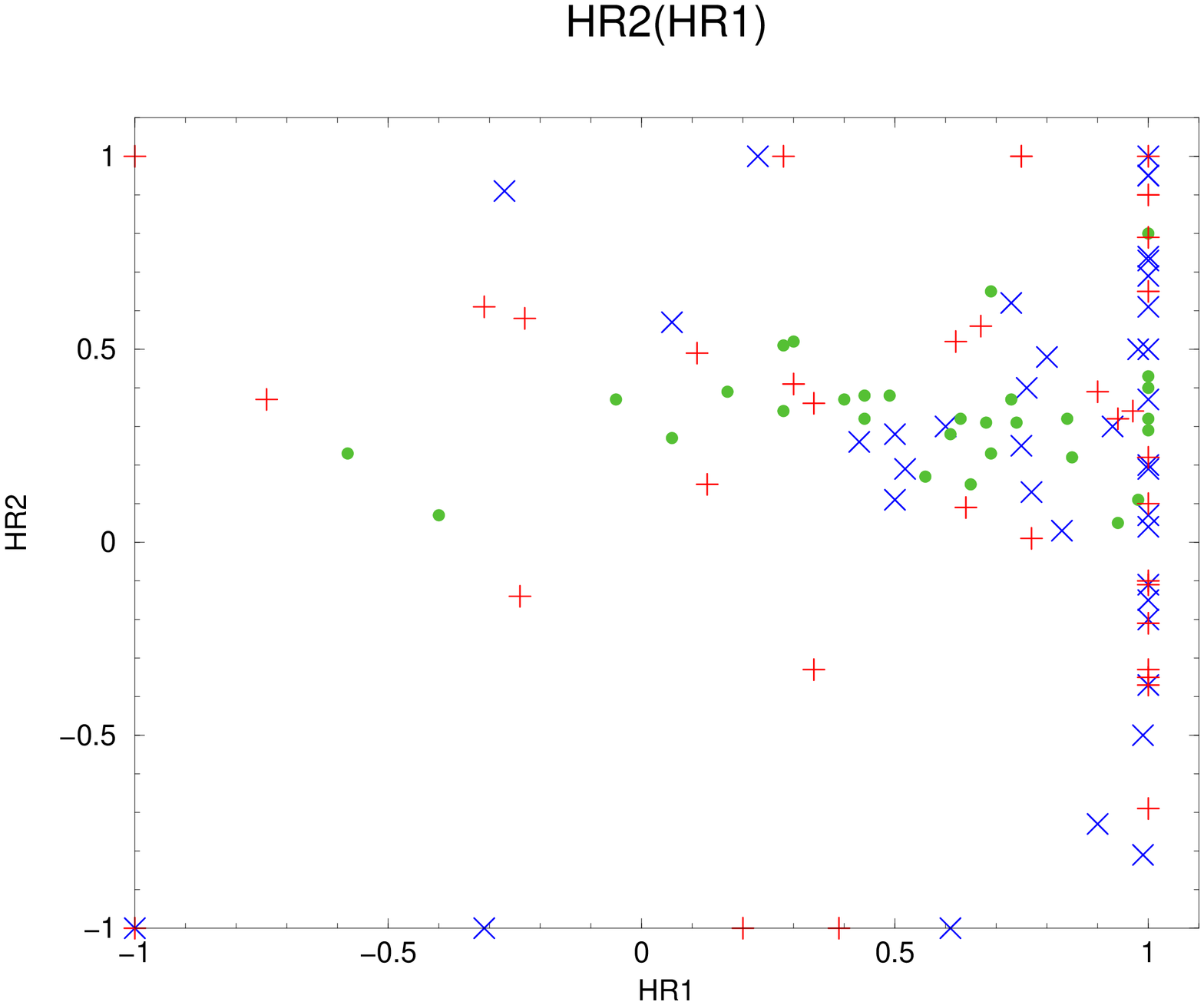,height=10cm,angle=-90}}
\caption{Top --- X--ray hardness ratios HR1 and HR2 for known AGNs among
our sources (BSC only). Bottom~--- the ratios of our sources 
(BSC\,+\,FSC), blue crosses ($\times$)~-- 1st class,
red crosses (+)~-- 2nd class and
points~-- 3rd class of the sources}
\end{figure}

A visual inspection of all 142 BSC\,+\,FSC sources in spiral and irregular 
galaxies has been done. These sources have been classified in 4 groups:\\
1st class of clear offset (non--nuclear) sources, without probable 
contamination with a star--like object in X--ray error boxes~--- 37 sources;\\
2nd class of probable offset, but may be nuclear sources~--- 43 sources;\\
3rd class of very probable nuclear sources, unknown AGNs can be present
in this class,~--- 35 sources;\\
4th class of probable contaminations (star--like objects inside a box, or 
unusually big distance between X--ray source and galaxy.

Fig.\,4 presents X--ray hardness ratios HR1 and HR2 for known AGNs among
our sources (from BSC only) and for our sources from BSC\,+\,FSC.
There are many X--ray sources with hard spectra among our best correlations.
The known X--ray binaries in ROSAT data show very hard spectra (Motch
et at., 1998). This result is in agreement with interpertation of
the sources as X--ray binaries.

Fig.\,5 shows the hardness ratios HR1 and HR2 versus X--ray luminosity 
for 3 our classes of sources. The only firm conclusion may be done here,
that nuclear sources are brighter. They include really unknowm AGNs.

\begin{figure}
\vspace{-3cm}
\centerline{\psfig{figure=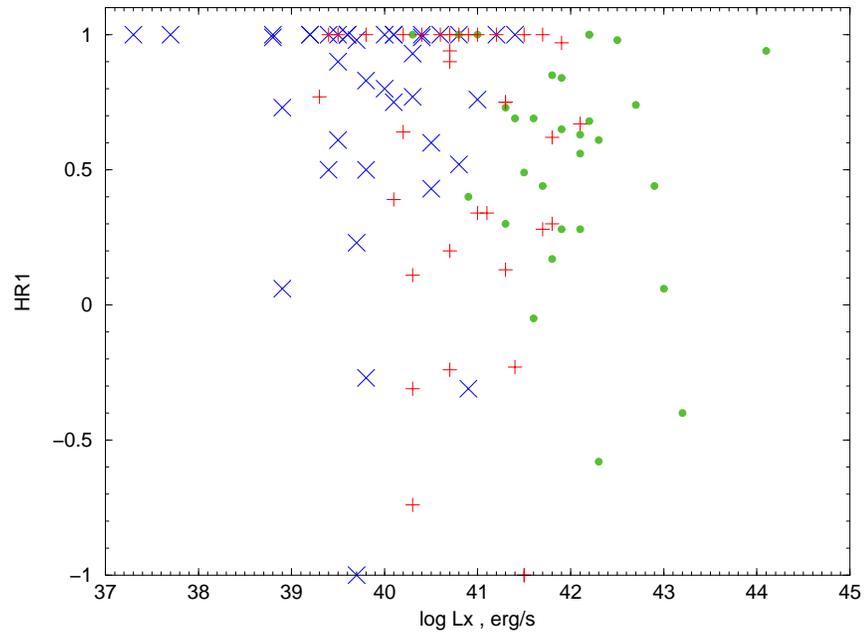,height=10cm,angle=-90}}
\centerline{\psfig{figure=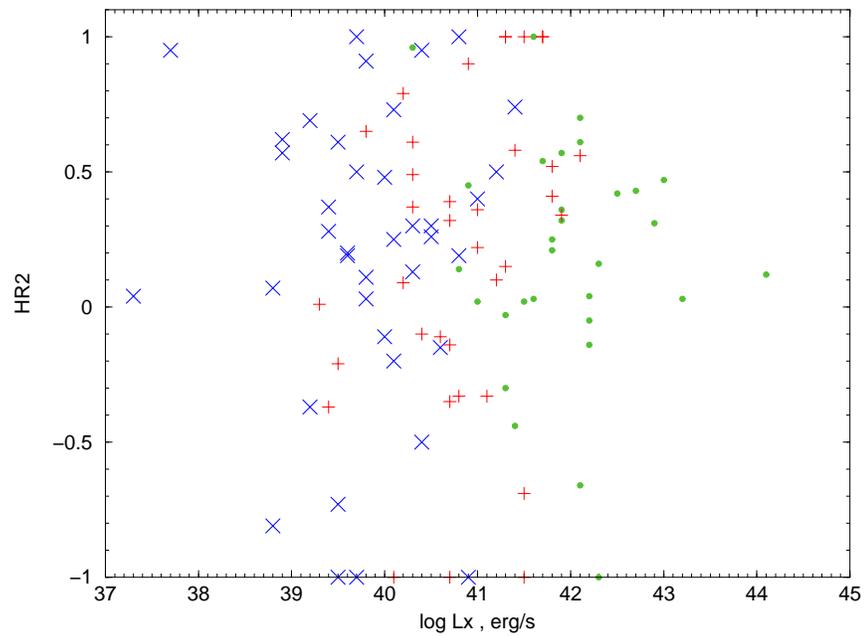,height=10cm,angle=-90}}
\caption{The hardness ratios versus X--ray luminosity 
for 3 our classes of sources: blue crosses ($\times$)~--
1st class, red crosses (+)~-- 2nd class and points~-- 3rd class}
\end{figure}

\begin{figure}
\centerline{\psfig{figure=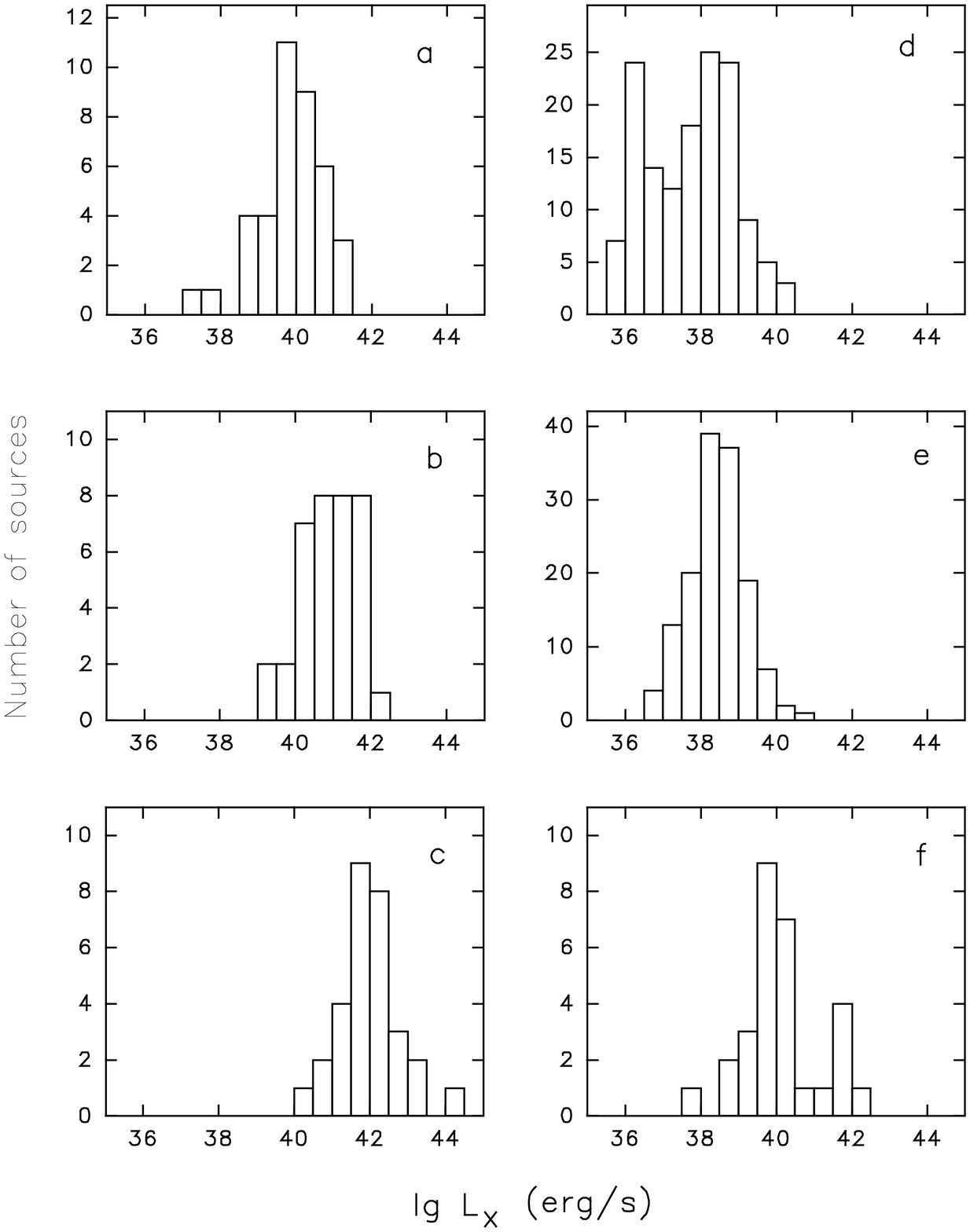,height=14cm}}
\caption{X--ray luminosities of our sources (a~-- 1st class, b~-- 2nd, 
c~-- 3rd) and galactic sources from Read, Pomnan and 
Strickland~(1996) (d), offset sources (e) and AGNs in nearby galaxies 
(f) from Roberts, Warwick~(2000)}
\end{figure}

In Fig.\,6 we present distributions of X--ray luminosities of our 
sources (a, b, c) and galactic sources from Read, Pomnan and 
Strickland~(1996) (d), offset sources (e) and AGNs (f) from 
Roberts, Warwick~(2000). The last is a study of complete list of nearby 
galaxies with the HRI of ROSAT. One can see again that our 2dn and 3rd
classes do include new AGNs.

\begin{figure}
\vspace{-3cm}
\centerline{\psfig{figure=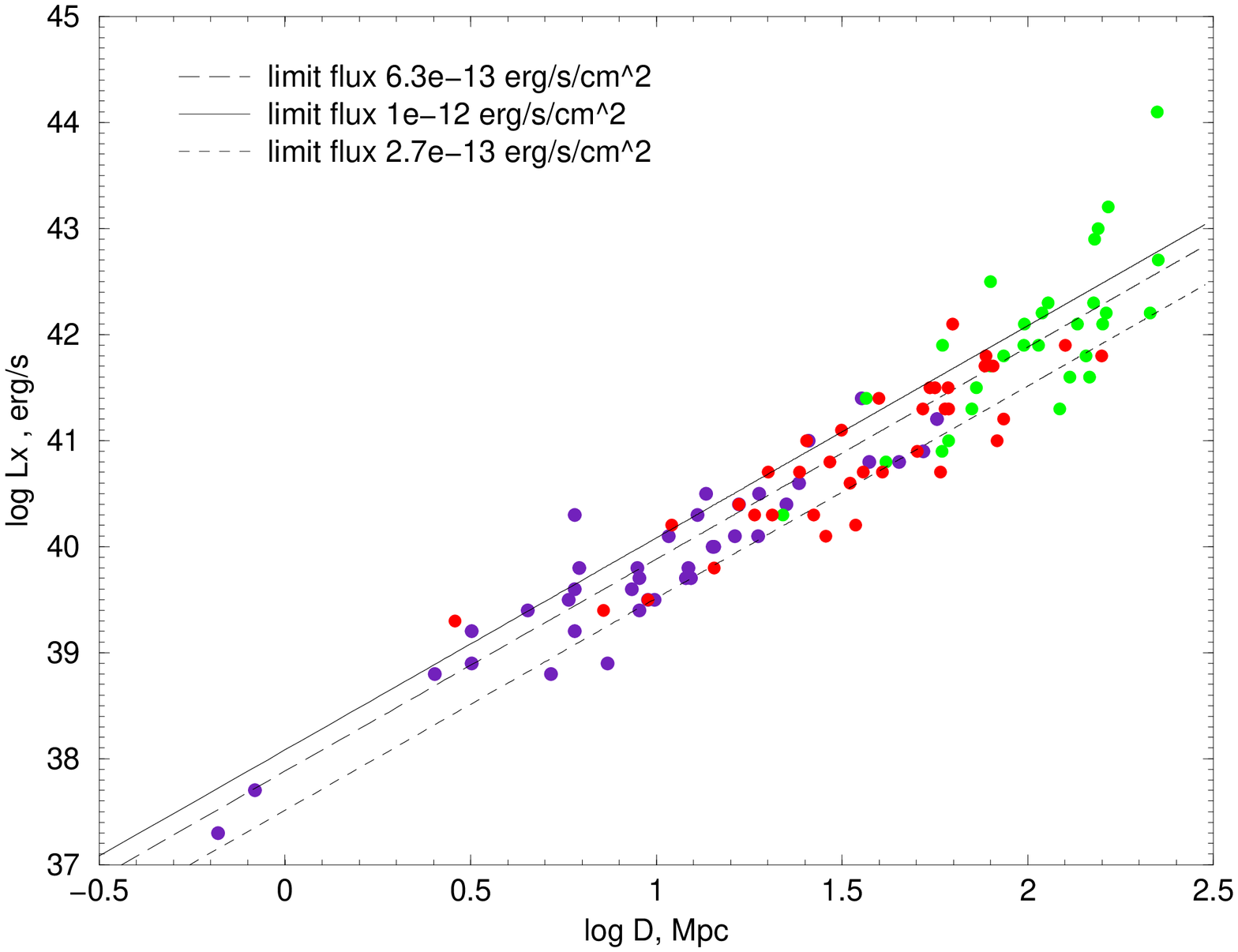,height=10cm,angle=-90}}
\caption{X--ray luminosities versus galactic distances (blue~-- 1st class,
red~-- 2nd and green~-- 3rd class) for positive correlations. The lines show
BSC and FSC limits}
\end{figure}

\begin{figure}
\centerline{\psfig{figure=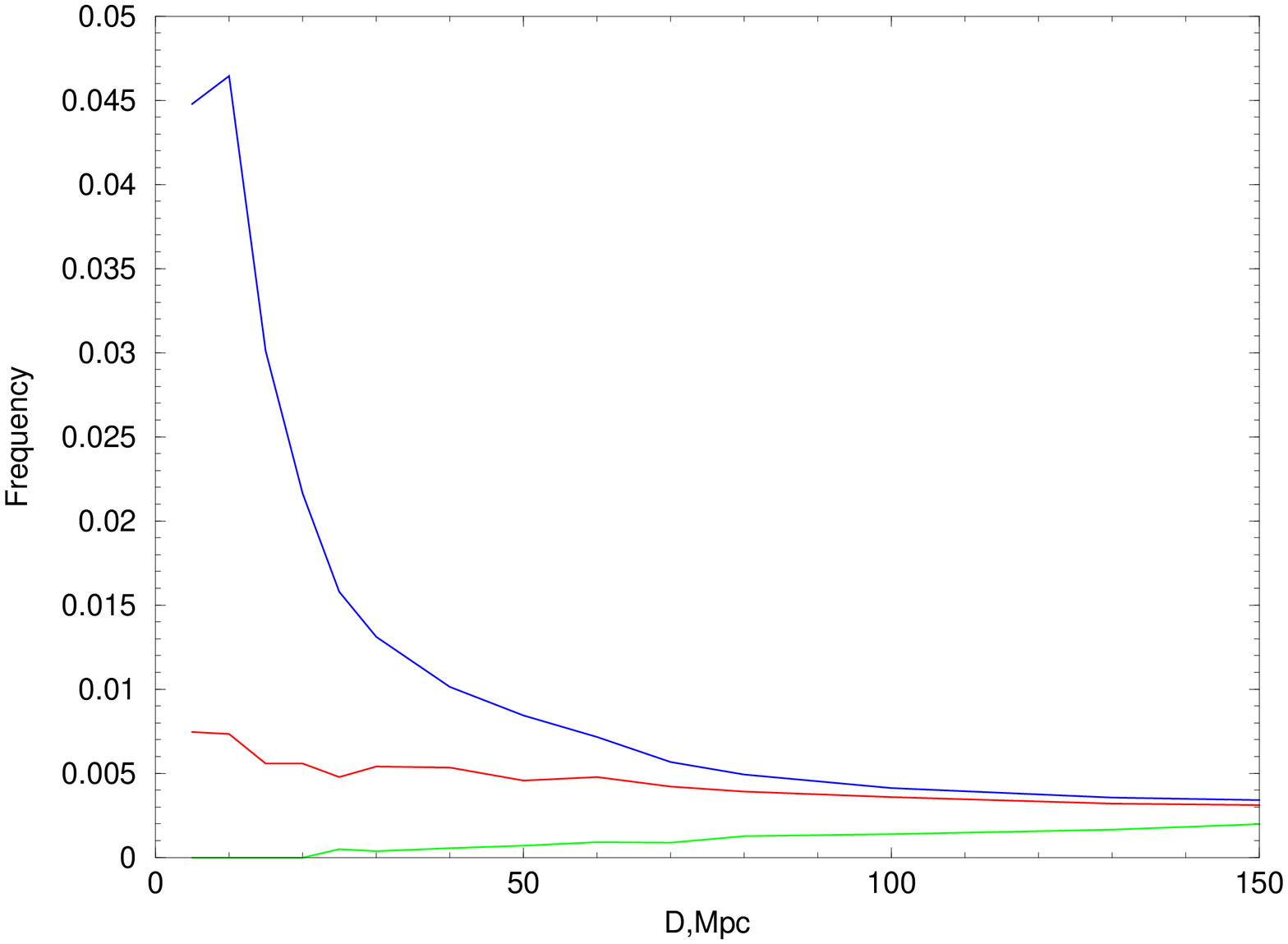,height=10cm,angle=-90}}
\caption{Frequencies of our positive correlations (blue~-- 1st class,
red~-- 2nd and green~-- 3rd class) per galaxy}
\end{figure}

We may compare our results with results
by Roberts and Warwick~(2000). Using only spiral galaxies in both samples,
taking their sources found with our ``limiting
flux'' $4 \, 10^{-13}$~erg/s and taking our sources found in their
``limiting galaxy magnitude'' $B \le 12\magdot5$ range of galaxies,
we find that they
found X--ray sources in 39~$\%$ of galaxies, but we found sources in
only $13~\%$ of galaxies. This contradiction may be because of observational
selection, Roberts and Warwick~(2000) used HRI data of pointing observations,
but we used ASS data, which were obtained without any preference of
astronomers. From the other hand a ratio of offset sources to AGNs found from
their flux--limited
data is 1.3, and the same ratio (we have not 3rd class sources in our
bright--galaxies sample, all they are known AGNs!) in our data 1.5, if
we consider the 1st + 2-nd class sources as non--nuclear over known AGNs.
And this ratio is 0.7, if we consider only 1st class sources as non--AGNs
over 2--nd class sources + known AGNs. We conclude that our
number of offset/nuclear sources
($0.7 \div 1.5$) is in agreement with the number obtainesd from  Roberts and
Warwick~(2000) data (1.3).

Fig.\,7 shows X--ray luminosities versus galactic distances for 
our three classes of positive correlations.
We see that the FSC may detect all
sources with $L_x > 10^{39}$~erg/s in galaxies closer that
4~Mpc, and all sources with  luminosities $L_x > 10^{40}$~erg/s
could be detected to a distance $D<12$~Mpc.

We calculated frequencies of the positive correlations in the
galaxies studied, that is a number of sources in S and Irr galaxies
over all galaxies of these types in the RC3
cataloque closer than some specific distance. In Fig.\,8 we observe
a high peak in the 1st class sources at distances $\approx 10$~Mpc.
Sources with luminosities $L_x = 10^{39 \div 40}$~erg/s contribute
mainly in the peak.
Otherwise at distances $> 10$~Mpc we have
a strong selection in our data and can not recognize all sources with
$L_x > 10^{40}$~erg/s. We may estimate a frequency using only closer
galaxies, in the interval of distances $D < 10$~Mpc.
The frequency of the best offset sources (the 1-st class)
is about $4 \div 5\,\%$ per S/Irr galaxy in this interval.
They are ultra--bright
X--ray sources, $L_x \sim 10^{39} \div 10^{41}$~erg/s.
Frequencies of sources of the 2-nd and 3-rd classes show quite different
behaviour. They confirm the interpretation, that in 2-nd class of ''probably
offset'' sources we have both stellar sources and AGNs, but among
the 3-rd class (the known AGNs being excluded) we have unknown AGNs.

We conclude that the data do not contradict
the idea of existence of the ``hypothetical'' face--on SS\,433
X--ray sources.
The only way to identify such stars is their expected violent 
variability. The problem of existence of ultrabright X--ray sources in
galaxies is well known (Fabbiano, 1998; Roberts and Warwick, 2000).
We may identify these sources, on the base of today--level
of knowledges, as face--on SS\,433 stars.

\end{document}